\documentclass[iop]{emulateapj}
\usepackage{amsmath,color}
\usepackage{array}
\usepackage{float}
\usepackage{graphicx}
\usepackage{subfigure}
\usepackage{color}

\usepackage{natbib}

\begin{document}
\title{Multiband photometry of a Patroclus-Menoetius mutual event: Constraints on surface heterogeneity}
\author{Ian Wong\altaffilmark{1,2,}\altaffilmark{$\dag$} and Michael E. Brown\altaffilmark{2}}
\affil{\textsuperscript{1}Department of Earth, Atmospheric, and Planetary Sciences, Massachusetts Institute of Technology,
Cambridge, MA 02139, USA; iwong@mit.edu\\
	\textsuperscript{2}Division of Geological and Planetary Sciences, California Institute of Technology,
Pasadena, CA 91125, USA \\
	\textsuperscript{$\dag$}51 Pegasi b Postdoctoral Fellow}

\begin{abstract}
We present the first complete multiband observations of a binary asteroid mutual event. We obtained high-cadence, high-signal-to-noise photometry of the UT 2018 April 9 inferior shadowing event in the Jupiter Trojan binary system Patroclus-Menoetius in four Sloan bands --- $g'$, $r'$, $i'$, and $z'$. We use an eclipse lightcurve model to fit for a precise mid-eclipse time and estimate the minimum separation of the two eclipsing components during the event. Our best-fit mid-eclipse time of $2458217.80943^{+0.00057}_{-0.00050}$ is 19~minutes later than the prediction of \citet{grundy}; the minimum separation between the center of Menoetius' shadow and the center of Patroclus is $72.5\pm0.7$~km --- slightly larger than the predicted 69.5~km. Using the derived lightcurves, we find no evidence for significant albedo variations or large-scale topographic features on the Earth-facing hemisphere and limb of Patroclus. We also apply the technique of eclipse mapping to place an upper bound of $\sim$0.15~mag on wide-scale surface color variability across Patroclus.

\end{abstract}

\section{Introduction}\label{sec:introduction}

The origin and nature of Jupiter Trojans have remained an enigma for many decades. The central question remains whether these objects orbiting in 1:1 mean motion resonance with Jupiter formed \textit{in situ} or were scattered inward from the outer Solar System and captured into resonance during a period of dynamical instability sometime after the end of planet formation \citep{gomes,morbidelli,tsiganis}. While recent numerical modeling has demonstrated the consistency of the latter scenario with current theories of late-stage giant planet migration \citep[e.g.,][]{roig2}, the definitive answer to the question of the Trojans' formation location will invariably come from obtaining a more detailed understanding of the physical properties and composition of these objects.

The discovery of Menoetius, the nearly equal-size binary companion of Patroclus \citep{merline}, established the first multiple system in the Trojan population and provided the first estimate of a Trojan's bulk density. Subsequent analyses using resolved imaging \citep{marchis,grundy}, thermal spectroscopy during mutual events \citep{mueller}, and stellar occultations \citep{buie} have refined the density estimate to the current value of $1.08\pm0.33$~g/cm$^{3}$. This low density indicates that Patroclus-Menoetius's bulk composition is dominated by ices, with significant porosity, similar to density measurements of cometary nuclei. Such a compositional model points strongly to an outer solar system origin of Trojans. 

Theories of binary asteroid formation center around two processes: capture or coeval formation. The former process involves stochastic close encounters, between two bodies, with capture occurring either via dynamical friction from surrounding objects, energy exchange during gravitational scattering of a third body, or capture of fragments from a collision \citep[e.g.,][]{goldreich}. Within the context of dynamical instability models of solar system evolution, Patroclus-Menoetius could have formed via capture early on during the planet formation stage, after the planet formation stage prior to the instability in the outer Solar System, or following the scattering of Trojans into their current orbits. The latter process of coeval formation forms binaries through the gravitational collapse of locally concentrated swarms of planetesimals \citep[e.g.,][]{nesvornycollapse}.

While coeval formation has a strong tendency to produce near-equal binary components, capture typically results in large size discrepancies between the two components. Therefore, the near-equal sizes of Patroclus and Menoetius point toward coeval formation. Furthermore, coeval formation always produces companions with identical compositions, while capture scenarios can yield heterogeneous pairs. Detailed study of Kuiper Belt binaries has revealed a preponderance of equal-color pairs, whereas the average system colors span the full range of colors seen in the overall population \citep{benecchi}. If recent dynamical instability models are true, and the Trojans were scattered into their current orbits from the outer Solar System, then one would expect Patroclus-Menoetius to also have identical colors as a result of coeval formation in the early Solar System.

Comparisons of the properties of the two binary components provide a powerful empirical test of binary formation theories. In particular, the measurement of discrepant physical properties between Patroclus and Menoetius would immediately rule out coeval formation. It has been hypothesized for over a decade that the Trojans are comprised of two color sub-populations with distinct photometric and spectroscopic characteristics \citep[e.g.,][]{roig,wong}, and within the framework of dynamical instability models, these two sub-populations formed in different regions of the outer protoplanetary disk \citep{wong3}. If Patroclus and Menoetius are found to belong to different sub-populations, then it means that the binary system formed via capture during or after the period of dynamical instability, when the two sub-populations first mixed.

The unique nature of the Patroclus-Menoetius system has made it a prime target for detailed study, and it is one of five Trojan asteroids that will be visited by the space probe \textit{Lucy}. An extensive effort has begun to better characterize the Trojan targets in order to maximize the mission's scientific yield. In 2017--2019, Patroclus-Menoetius was in a mutual event season when eclipse and occultation events were visible from Earth. We obtained multiband photometric observations of an inferior shadowing event as Menoetius' shadow passed across Patroclus on UT 2018 April 9. In this paper, we present high-cadence, high-signal-to-noise lightcurves in four bands and fit the eclipse lightcurves to produce a precise mid-eclipse timing and estimate of the relative separation of the eclipsing components at mid-eclipse. We also use the technique of eclipse mapping, a first in the study of binary asteroids, to derive constraints on surface heterogeneity from the resultant color lightcurves.

\section{Observations and Data Analysis}\label{sec:obs}

We observed the UT 2018 April 9 Patroclus-Menoetius inferior eclipsing event using the then newly-installed Wafer-scale Imager for Prime (WaSP) instrument on the 200-inch Hale Telescope at Palomar Observatory. The science detector in WaSP is a 6144$\times$6160 CCD with a pixel scale of 0.18$''$. We chose a 2048$\times$2048 sub-array to reduce readout time and increase the cadence of our observations. As the shadow of Menoetius passed across the surface of Patroclus, we imaged the system in four Sloan filters --- $g'$, $r'$, $i'$, and $z'$ --- with individual exposure times of 30, 20, 20, and 45~s, respectively, which yielded a target signal-to-noise of at least 100 in all bands. Filters were cycled in the order $g'$-$r'$-$i'$-$z'$, producing a uniform cadence of roughly 5.5 minutes in each band, after accounting for readout and filter changes. Bias frames and dome flats were acquired at the beginning of the night prior to science observations.

Observing conditions at Palomar ranged from average to poor throughout the night. The sky was mostly clear, with a few isolated bands of thin, high-altitude clouds passing through at various points during the night. The seeing was poorest at the beginning of the observations, prior to the start of eclipse; before UT 5:00, the typical seeing exceeded 1.6$''$, going as high as 2.1$''$ at times. The remainder of the night saw significantly better seeing, averaging around 1.2-1.3$''$, with the exception of a roughly 30-minute period around UT 8:00, when there was a spike in the seeing to over 1.6$''$, likely associated with the passage of a few tenuous bands of high-altitude clouds across the vicinity of the observing field. There was also an increase in the seeing during the final 45 minutes of observation. These periods of relatively poor seeing can be identified by the corresponding notable increase in scatter in the lightcurves during those times.

\begin{figure}[t!]
\begin{center}
\includegraphics[width=\linewidth]{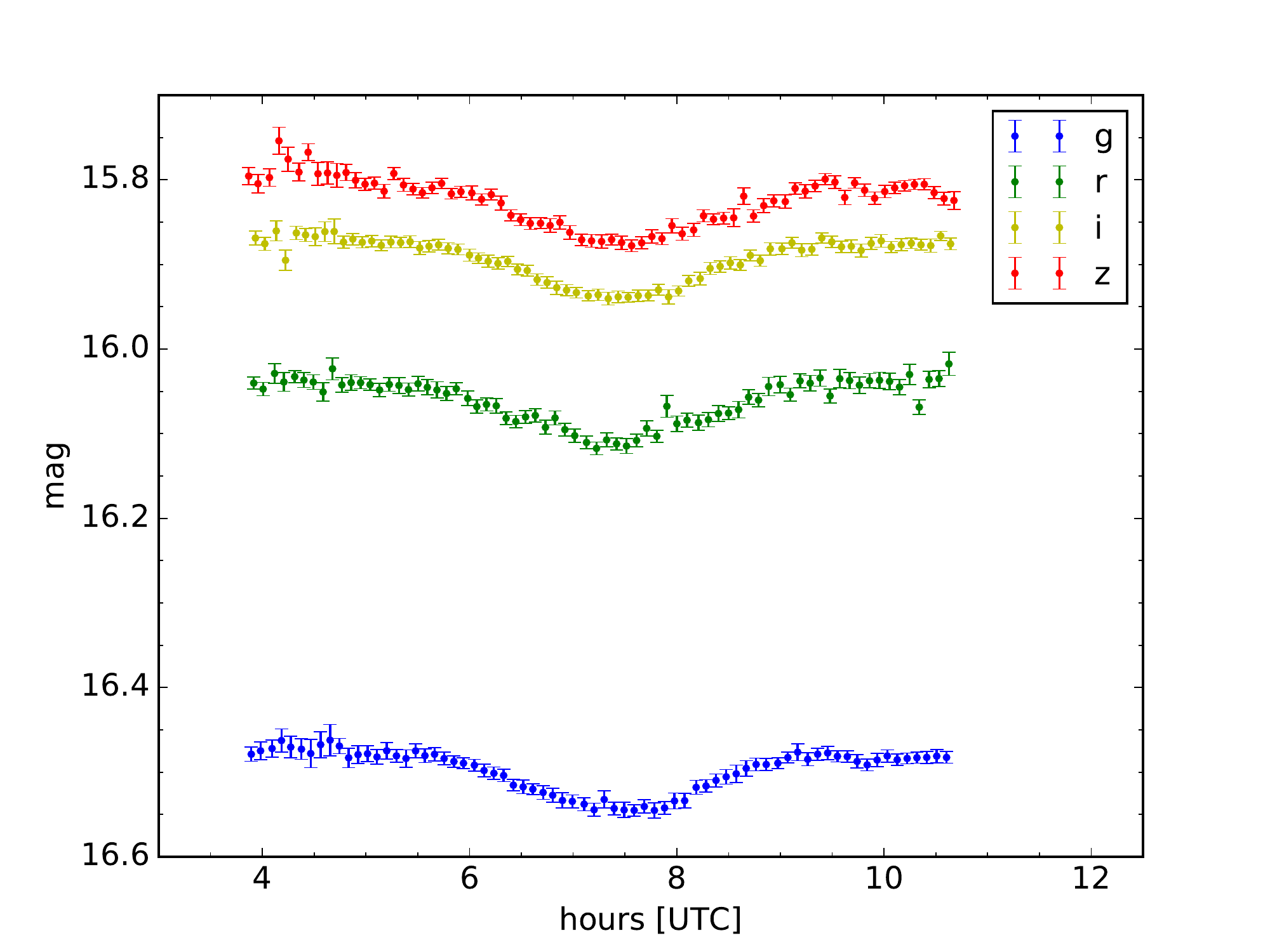}
\end{center}
\caption{Apparent magnitude lightcurves of the Patroclus-Menoetius system prior to, during, and following the inferior eclipsing event in the Sloan $g'$, $r'$, $i'$, and $z'$ bands. The vertical axis denotes increasing brightness (decreasing magnitude). Periods of larger scatter correspond to times of poorer observing conditions and higher seeing. The overall increased scatter in the $z'$-band lightcurve is attributed to discernible residual fringing on the images.} \label{lightcurves}
\end{figure}

Image processing and photometric calibration were carried out using standard techniques. After the images were bias-subtracted and flat-fielded, the centroid positions and fluxes of bright sources in each image were obtained using SExtractor \citep{sextractor}. These sources were then matched with stars in the Pan-STARRS DR1 catalog \citep{flewelling} to produce an astrometric solution and a photometric zeropoint. Our pipeline then automatically queried the JPL Horizons database for the position of Patroclus-Menoetius at the time of the exposure, identified the corresponding source on the image, and computed its apparent magnitude. Photometric extraction was carried out using a variety of fixed circular aperture sizes with diameters ranging from 8 to 24 pixels, choosing the optimal aperture for each exposure that minimizes the resultant photometric error. The median optimal aperture diameters in the four bands are 20, 11, 16, and 17 pixels, corresponding to radii of  $1.80''$, $0.99''$, $1.44''$, and $1.53''$, respectively.

In Figure~\ref{lightcurves}, the apparent magnitude lightcurves are plotted in each band; the individual 1$\sigma$ uncertainties are a quadrature sum of the propagated photometric errors stemming from the measured fluxes and the zeropoint uncertainties. The eclipse produced a roughly 0.15~mag dimming of the total system brightness in each of the four bands. The median photometric uncertainties are 0.0079, 0.0085, 0.0067, and 0.0074~mag in $g'$-, $r'$-, $i'$-, and $z'$-band, respectively. A handful of outliers are discernible, for example, two in the $r'$-band lightcurve at around UT 8:00 and 10:20. Visual inspection of these images did not reveal cosmic rays or any obvious chip artifacts that could have affected these points. By changing the extraction aperture used for those exposures, we found that the saliency of these outliers showed notable variation, suggesting a non-astrophysical cause. We also note that all of the outlier exposures occurred during the periods of increased seeing mentioned previously. We have chosen to leave them in the lightcurves presented in this paper.

In $z'$-band images, there was discernible residual fringing on the flux arrays, even after flat-fielding, particularly in the northeast corner. While the target mostly avoided the regions of the detector with the most severe residual fringing, there is still a noticeable effect in the $z'$-band lightcurve, as manifested by the larger scatter in the photometry on short timescales and larger than expected photometric zeropoint errors. We do not attempt to correct for fringing, and while we present the $z'$-band lightcurve in Figure~\ref{lightcurves}, we do not utilize or discuss the $z'$-band photometry in the following analysis.

\section{Discussion}\label{sec:discussion}

\subsection{Eclipse lightcurve fit}\label{subsec:model}

To derive estimates of the mid-eclipse time and the extent of the eclipsed region, we use a custom transit model to fit the $i'$-band lightcurve, which has the smallest median photometric error. Since the eclipsed region of Patroclus is non-illuminated, we can equivalently model the eclipse event as an occultation. 

The mutual orbit of the binary system is consistent with circular, so we fix the eccentricity to zero. We fix the orbital period and semimajor axis to the values reported and assumed in the mutual event predictions of \citet{grundy}: $P=4.282680$~days, $a=688.5$~km. Both components are significantly non-spherical, and modeling from occultation and rotational phase curves yields a triaxial radius ratio of $\alpha:\beta:\gamma = 1.3:1.21:1$; the long dimension of each object lies along the line connecting the two objects, while the shortest dimension is aligned with the angular momentum vector of the binary system \citep{buie}. During a mutual event, the sky-projected shapes of Patroclus (1) and Menoetius (2) are ellipses with semimajor axis values of $\beta_1=117$~km, $\gamma_1=98$~km and $\beta_2=108$~km, $\gamma_2=90$~km, respectively. 

We fit for the center of eclipse time $T_{c}$ and the apparent orbital inclination $i$, which is defined relative to the sky plane so that $i=90^{\circ}$ is a perfectly edge-on occultation where the centers of the two objects align at mid-event. For each pair of $T_{c}$ and $i$ values in the Markov Chain Monte Carlo (MCMC) chain, we use the orbital shape and period to derive the relative separation vector between the two components at every point in the time series. To compute the amount of Patroclus blocked by Menoetius' shadow, we use a Python-based code\footnote{https://github.com/chraibi/EEOver} to calculate the overlapping area of the two ellipses, which is based on the algorithm described in \citet{ellipse}. We also fit for a constant multiplicative factor to normalize the out-of-eclipse lightcurve to unity.

\begin{figure}[t!]
\begin{center}
\includegraphics[width=\linewidth]{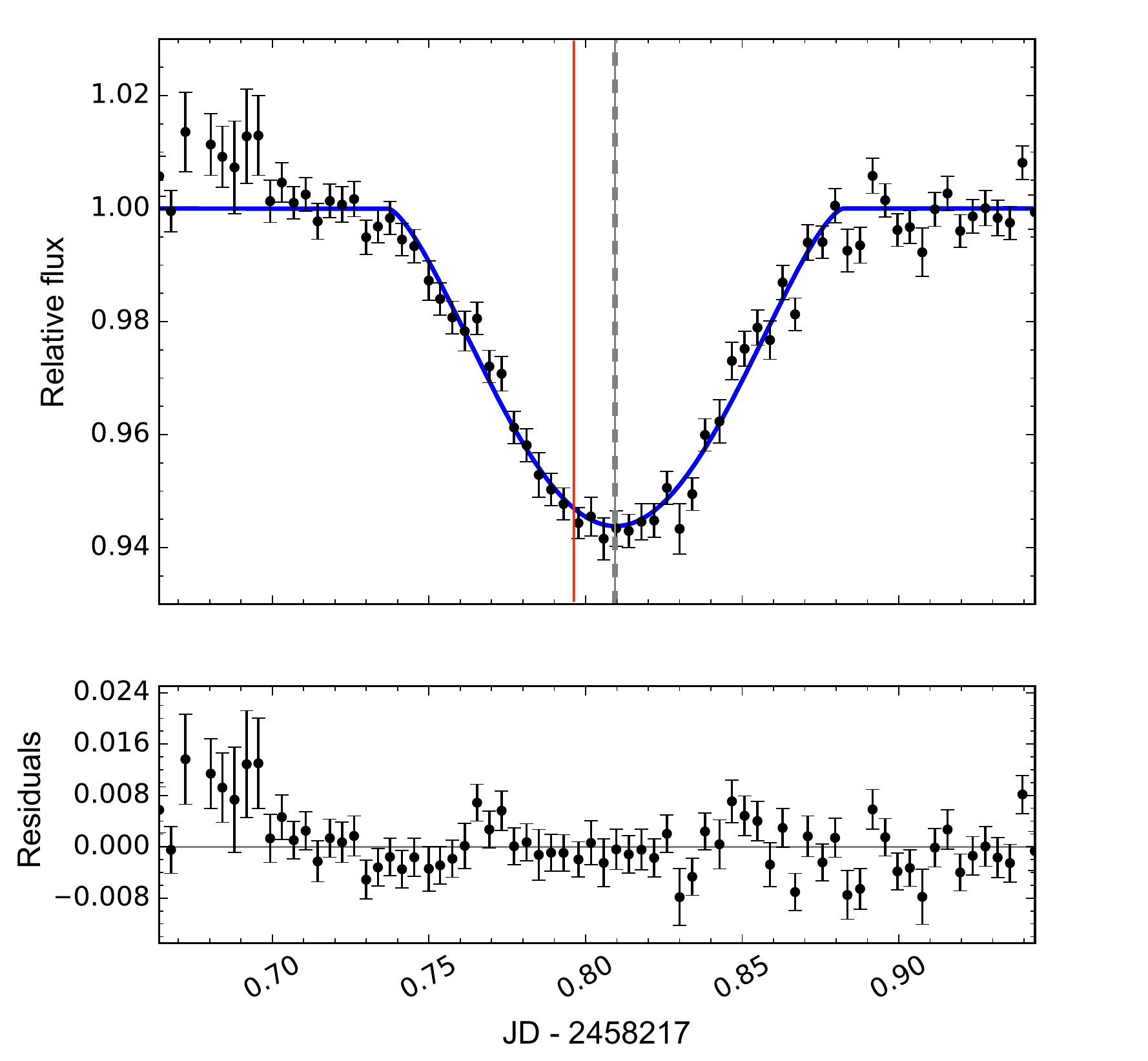}
\end{center}
\caption{Top panel: $i'$-band lightcurve of the mutual event (blacks points) along with the best-fit eclipse lightcurve model (blue line). The out-of-eclipse combined brightness of Patroclus and Menoetius is normalized to unity. Vertical black lines indicate the best-fit mid-eclipse time and uncertainties: $T_{c} = 2458217.80943^{+0.00057}_{-0.00050}$; the red line shows the mid-eclipse time predicted by \citet{grundy}: 2458217.7965. Bottom panel: corresponding residuals from the fit. The scatter in the residuals is 0.0048, while the median per-point flux uncertainty in 0.0030.} \label{fit}
\end{figure}

We modify the transit model to account for the fact that Menoetius is illuminated, which dilutes the transit signal relative to the case where the secondary is dark. If the lightcurve of the eclipsed object Patroclus is modeled as $\lambda(t)$, then the total lightcurve of the binary system is $(\lambda(t)+f_{2})/(1+f_{2})$, where $f_{2}$ is the brightness of the secondary Menoetius relative to Patroclus. If Patroclus and Menoetius were identical in albedo, then the brightness ratio would be equal to the ratio in sky-projected areas: $f_{2}=\beta_2\gamma_2/\beta_1\gamma_1$. While it is reasonable to assume that the two components are largely identical in composition and therefore should have very similar albedos, given the likely formation mechanism of such near-equal mass binaries (see Section~\ref{sec:introduction}) and the markedly narrow albedo distribution of the Trojan asteroid population as a whole \citep[e.g.,][]{romanishin,fernandez}, we nevertheless account for our uncertainty in the albedos of the individual components: we set a multiplicative scaling factor on $f_{2}$ and place a Gaussian prior on its value centered on unity with a standard deviation of 20\%, consistent to the variance in the measured geometric albedos of large Trojans \citep[e.g.,][]{fernandez,romanishin}.

The best-fit eclipse lightcurve is plotted in Figure~\ref{fit}. We have removed the fourth data point prior to the final fit, which is more than $3\sigma$ discrepant from the best-fit eclipse model. The lightcurve is normalized such that the combined out-of-eclipse brightness of Patroclus and Menoetius is unity. The scatter in the residuals is 0.0048, compared to a median relative flux uncertainty of 0.0030, indicating significant non-white noise in the lightcurve attributable to the periods of poorer observing conditions at the beginning and towards the end of the night. We measure a mid-eclipse time (in Julian days) of
\begin{equation}T_{c} = 2458217.80943^{+0.00057}_{-0.00050},\end{equation}
which corresponds to UT 2018 April 9 7:25:35 with an uncertainty of 46s. This is 19~minutes later than the predicted center of eclipse in \citet{grundy}. 

Meanwhile, we obtain a precise relative inclination estimate of $i=83.95\pm0.06$~deg. We can compute the sky-projected separation $d_{\mathrm{min}}$ of the center of Patroclus and the center of Menoetius's shadow at mid-eclipse:
\begin{equation}d_{\mathrm{min}} = a\cos(i) = 72.5\pm0.7~\mathrm{km}. \end{equation}
\citet{grundy} reported a predicted minimum separation between the centers of the two eclipsing bodies of 69.5~km. The greater separation derived from our fit indicates a more grazing shadowing event than predicted and points toward a slight inaccuracy in the orbital pole obliquity calculated in \citet{grundy}. We remind the reader that during this event, it is the shadow of Menoetius that occults Patroclus. The disk of Menoetius itself does not interact with the disk of the primary.

\subsection{Surface properties}\label{subsec:color}

Various physical and compositional properties of the surface are expressed in the eclipse lightcurves. When looking in one photometric band, comparison between the observed lightcurve and the best-fit eclipse model provides constraints on albedo variations across the eclipsed region of the primary as well as the shapes of both binary components. Significant covariant deviations in the residuals from a flat line may indicate patches of enhanced or reduced reflectivity on the primary or significant deviations along the limb from that of a sky-projected ellipse. Examining the residuals from our best-fit eclipse model in Figure~\ref{fit}, we do not discern any statistically significant deviations indicating non-uniform reflectivity or non-ellipsoidal shapes for the primary disk and secondary shadow.

\begin{figure}[t!]
\begin{center}
\includegraphics[width=\linewidth]{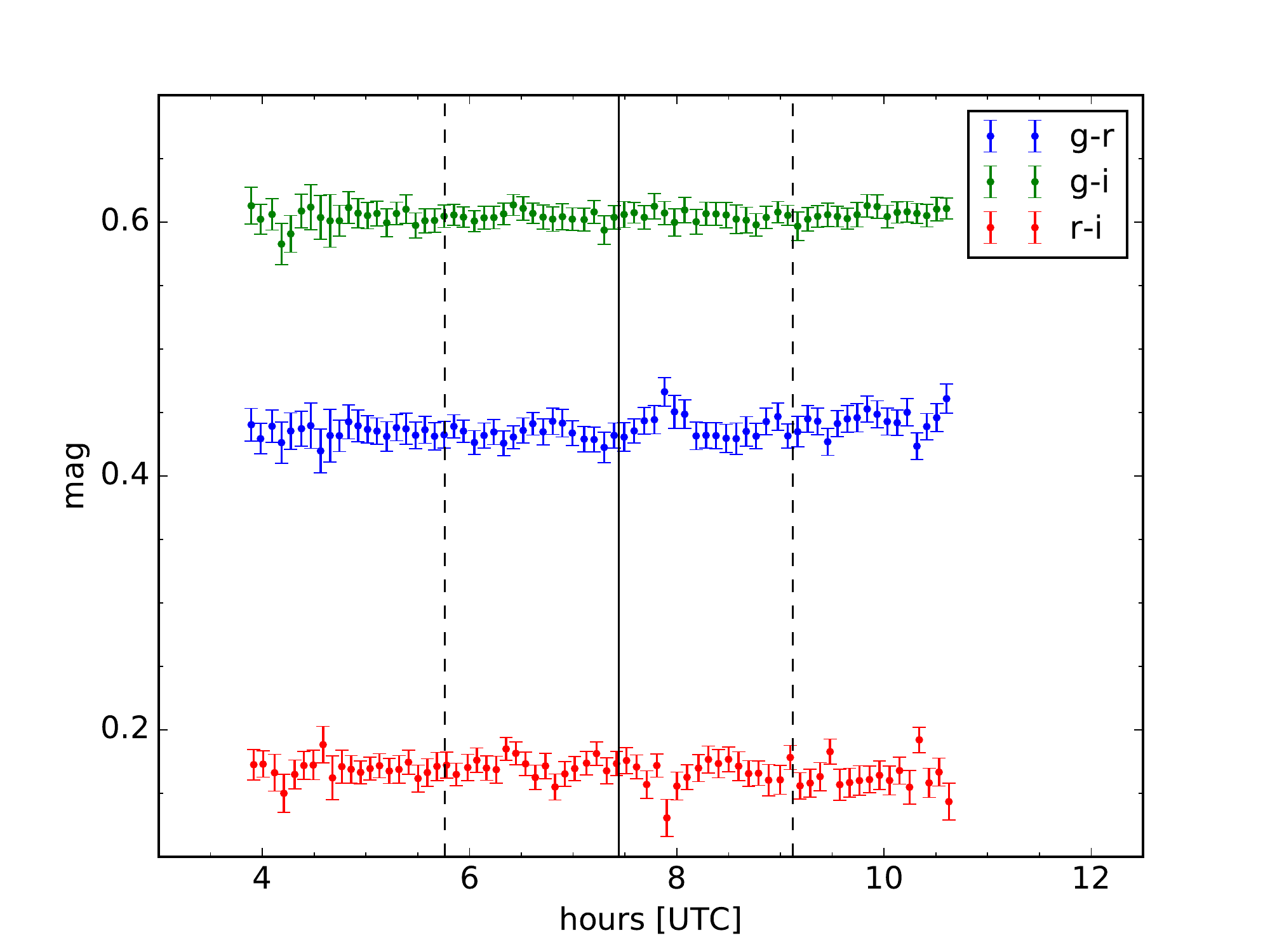}
\end{center}
\caption{Color lightcurves derived from the photometric lightcurves in Figure~\ref{lightcurves}, showing minimal variations during the shadowing event. The vertical solid and dashed lines indicate mid-eclipse and the beginning/end of the eclipse event, respectively. Almost all points in the color lightcurves are consistent with a flat line to within $1.5\sigma$. The two notable outliers at around UT 8:00 and 10:20 in the $g-r$ and $r-i$ lightcurves stem from two outlier points in the $r'$-band lightcurve (see Figure~\ref{lightcurves}).} \label{color}
\end{figure}

Leveraging photometric lightcurves at multiple wavelengths provides additional information about the level of color variation across and between the two binary components. As the shadow of Menoetius eclipses Patroclus, the contribution of the shadowed region to the average color of the system is removed. By examining the resultant color lightcurves, one can piece together the color distribution of the eclipsed region in a technique known as eclipse mapping. This powerful method allows one to potentially extract spatial information about the target from spatially unresolved images. For each pair of photometric lightcurves, we use linear interpolation between adjacent points in the second lightcurve's time series to calculate the magnitudes in the second filter at the time sampling of the first lightcurve's time series. We then subtract the resampled lightcurves from one another, adding the propagated uncertainties in quadrature. Figure~\ref{color} shows the three color lightcurves derived from the $g'$-, $r'$-, and $i'$-band lightcurves in Figure~\ref{lightcurves}. We have omitted the color lightcurves involving $z'$-band due to the effect of residual fringing (see Section~\ref{sec:obs}).

The color lightcurves are generally very smooth, with no large deviations and almost all points lying well within $1.5\sigma$ of the average color across the observations. We note that the regions with increased short-term variation and the largest color deviations correspond precisely to the periods during our observations when seeing was poor and highly variable (see Section~\ref{sec:obs}). Given the grazing nature of this eclipse event, we are only sensitive to very large color variations on small scales. The most stringent constraints on color variability can be derived from comparing the mid-eclipse color, when the eclipsed region is at its maximum, with the out-of-eclipse color. For all color lightcurves, the mid-eclipse color value is well within $1\sigma$ of the out-of-eclipse color, so we place $1\sigma$ upper bounds on the color variability using the median color uncertainty from the lightcurves, $\sigma_{c}$.

To quantify these constraints, we consider two cases. The first case seeks to constrain the difference between the average color of the eclipsed region on Patroclus $c^*$ and the average color $c$ of the uneclipsed regions on both objects. The change in the measured color of the combined system between the out-of-eclipse baseline and mid-eclipse is weighted by the ratio of the maximum eclipsed area $A^*$ to the uneclipsed area $A_{1}+A_{2}-A^*$, where $A_{1}=\pi\beta_{1}\gamma_{1}$ and $A_{2}=\pi\beta_{2}\gamma_{2}$ are the sky-projected areas of Patroclus and Menoetius, respectively. The maximum eclipsed area of Patroclus, as derived from our eclipse model fit in Section~\ref{subsec:model}, was $12.4\%$ of its sky-projected disk: $A^*=1110~\mathrm{km}^2$. From here, the difference in color $\Delta c_{1} \equiv |c^*-c|$ is given by
\begin{equation} \Delta c_{1} = \sigma_{c}\frac{A_{1}+A_{2}-A^*}{A^*}.\end{equation}
For the $g-i$ color variability, for example, we have $\sigma_{g-i}=0.0092$ and establish an upper limit of  $\Delta c_{1,g-i} = 0.13$~mag, with similar constraints for the other colors.

The second case assumes that the two components have different colors, $c_{1}$ and $c_{2}$, but are individually uniform in color. A similar derivation yields the following expression for $\Delta c_{2} \equiv |c_{2}-c_{1}|$:
\begin{equation}\Delta c_{2} = \sigma_{c}\frac{(A_{1}+A_{2}-A^*)(A_{1}+A_{2})}{A^*A_{2}}.\end{equation}
The constraints on $\Delta c_{2}$ are much looser. For $g-i$, this upper limit is $\Delta c_{2,g-i} = 0.28$~mag.

Starting with the second constraint, we see that the small maximum shadow coverage of Patroclus prevents us from deriving particularly useful upper limits on the difference in color between the two components. For comparison, the two color sub-populations in the Trojans have mean $g-i$ colors of 0.73 and 0.86 \citep{wong,wong2}, so a larger eclipsed area and/or more precise photometry would be needed to confidently rule out a binary comprised of components from two different sub-populations using lightcurves like these. Typical color differences between the components of KBO binary systems are also significantly smaller than our upper bound constraint \citep[e.g.,][]{benecchi}.

The first constraint reflects the level of large-scale surface inhomogeneities across Patroclus. This much more stringent constraint suggests that the surface of Patroclus is quite homogeneous. When comparing with other ice-rich asteroids and satellites that have well-mapped surface color distributions, we find that those larger bodies, such as Pluto, Europa, Ceres, and Triton, have significantly higher levels of color variability than Patroclus across physical scales comparable to the relative area probed by our eclipse measurements. In addition, those objects also display significant localized albedo variations across the surface, which we do not detect on Patroclus from our measurements.

The relative homogeneity of Patroclus is consistent with theories regarding the formation and evolution of Trojans and similar objects. Whereas the larger bodies like the Galilean satellites and dwarf planets accreted sufficient material to gravitationally circularize, internally differentiate, and, in some cases, bind tenuous atmospheres, leading to secondary geological processes that continue to be active in the present day, smaller bodies like the Trojans would have formed as undifferentiated ice-rock agglomerations, similar to cometary nuclei, without sufficient gravity or internal heating to undergo further physical or compositional alterations \citep[e.g.,][]{wong3}. These primitive objects would have a uniform composition throughout and develop a homogeneous irradiation mantle across their entire surfaces.

Such a formation scenario does not preclude occasional instances of surface inhomogeneities due to minor cratering events. Areas of pristine material excavated by impacts might have much higher albedo than the $\sim$5\% typical of Trojans \citep[e.g.,][]{fernandez}. Likewise, these newly-exposed regions might have a distinct color from the rest of the radiation-reddened surface \citep{wong3}. Both the reflectivity and color inhomogeneities would be detectable using high-precision multiband lightcurves of mutual events similar to the ones presented in this work.

\section{Summary}

In this paper, we presented multiband photometric observations of the UT 2018 April 9 inferior shadowing in the Patroclus-Menoetius system. Our short-cadence high-signal-to-noise lightcurves provided a precise mid-eclipse timing measurement, $T_{c} = 2458217.80943^{+0.00057}_{-0.00050}$, which is later than the prediction from \citet{grundy} by almost 20 minutes. Eclipse lightcurve modeling showed that the eclipse magnitude was slightly less than predicted, with a minimum separation distance of $72.5\pm0.7$~km between the centers of Patroclus and Menoetius' shadow at mid-eclipse. Through an analysis of the color trends derived from the photometric lightcurves, we placed a moderately tight upper bound on the level of surface variability across Patroclus, in agreement with the predictions from formation models of primitive icy bodies. Meanwhile, the grazing nature of the event prevented us from ruling out a mixed binary scenario with components from different color sub-populations. Nevertheless, our analysis demonstrated the applicability of the eclipse mapping technique to the study of binary asteroids. Future work combining the observations of Patroclus-Menoetius from the 2017--2019 mutual event season with previous measurements will greatly improve the orbital parameters of the system. New orbital fits and shape models will enable more detailed planning of the \textit{Lucy} flyby encounter of the Patroclus-Menoetius system in 2033.

\acknowledgements

I.W. is supported by a Heising-Simons Foundation \textit{51 Pegasi b} postdoctoral fellowship. The Pan-STARRS1 Surveys (PS1) and the PS1 public science archive have been made possible through contributions by the Institute for Astronomy, the University of Hawaii, the Pan-STARRS Project Office, the Max-Planck Society and its participating institutes, the Max Planck Institute for Astronomy, Heidelberg and the Max Planck Institute for Extraterrestrial Physics, Garching, The Johns Hopkins University, Durham University, the University of Edinburgh, the Queen's University Belfast, the Harvard-Smithsonian Center for Astrophysics, the Las Cumbres Observatory Global Telescope Network Incorporated, the National Central University of Taiwan, the Space Telescope Science Institute, the National Aeronautics and Space Administration under Grant No. NNX08AR22G issued through the Planetary Science Division of the NASA Science Mission Directorate, the National Science Foundation Grant No. AST-1238877, the University of Maryland, Eotvos Lorand University (ELTE), the Los Alamos National Laboratory, and the Gordon and Betty Moore Foundation. This work made use of the JPL Solar System Dynamics high-precision ephemerides through the HORIZONS system.

\end{document}